\documentclass[epjST]{svjour}
\usepackage[T1]{fontenc}
\usepackage[ansinew]{inputenc}
\usepackage{graphicx}

\usepackage{amsmath,amssymb}
\usepackage{bm}
\usepackage{tabularx}
\usepackage{subfigure}
\usepackage[dvips]{color}

\def\be{\begin{equation}}
\def\ee{\end{equation}}
\def\ba{\begin{array}}
\def\ea{\end{array}}
\def\bea{\begin{eqnarray}}
\def\eea{\end{eqnarray}}

\def\Lap{\mathrm{Lap}}
\newcommand{\revision}[1]{#1}

\begin{document}
\title{Cracks in random brittle solids:}
\subtitle{From fiber bundles to continuum mechanics}
\author{
Sylvain Patinet\inst{1}\fnmsep\thanks{\email{sylvain.patinet@espci.fr}}
\and
Damien Vandembroucq\inst{1}\fnmsep\thanks{\email{damien.vandembroucq@espci.fr}}
\and
Alex Hansen,\inst{2}\fnmsep\thanks{\email{alex.hansen@ntnu.no}}
\and
St\'{e}phane Roux\inst{3}\fnmsep\thanks{\email{stephane.roux@lmt.ens-cachan.fr}}
}
\institute{
Laboratoire PMMH, ESPCI/CNRS-UMR 7636/Univ.
Paris 6 UPMC/Univ. Paris 7 Diderot, 10 rue Vauquelin,
F-75231 Paris cedex 05, France
\and
Institutt for fysikk, NTNU, N-7491 Trondheim, Norway
\and
LMT-Cachan, ENS-Cachan/CNRS/PRES UniverSud Paris, 61 Av.
Pr\'{e}sident Wilson, F-94235 Cachan cedex, France
}
\abstract{Statistical models are essential to get a better
  understanding of the role of disorder in brittle disordered solids.
  Fiber bundle models play a special role as a paradigm, with a very
  good balance of simplicity and non-trivial effects. We
  introduce here a variant of the fiber bundle model where the
  load is transferred among the fibers through a very compliant
  membrane. This Soft Membrane fiber bundle mode reduces to the
    classical Local Load Sharing fiber bundle model in 1D.
  Highlighting the continuum limit of the model allows to compute
  an equivalent toughness for the fiber bundle and hence discuss
  nucleation of a critical defect. The computation of the toughness
  allows for drawing a simple connection with crack front propagation
  (depinning) models.}  \maketitle
%


\section{Introduction}
\label{Sec:intro}

It is needless to emphasize the importance of brittle fracture for safety
and reliability issues.  In fact, the condition for crack {\em propagation}
is well understood since Griffith pioneering work\cite{Griffith}.
Critical energy release rates or equivalently stress intensity factors
are known to be material characteristics defining the onset of crack
propagation for an existing crack.  However, a resisting issue
concerns crack {\em initiation}.  Indeed, for a long period, the
common view was that cracks would initiate from preexisting defects.
Brittleness was invoked to consider that as soon as initiated a crack
would propagate in an unstable fashion down to complete failure.  Thus,
amazingly, solid mechanics for the most part disappeared from strength
predictions.  Because the initiation defect was local, the failure load
remained a statistical issue, with probability distribution which simply
resulted from that of the defect distribution after an elementary elastic
computation (together with some assumptions for the defect geometry).  This
was formalized within ``weakest links''  type of approaches, that can be
traced back to the pioneering work of Weibull.\cite{Weibull} These
approaches had also the merit of shedding some light on salient features
unexpected in the mechanics of homogeneous materials, such as the
occurrence of systematic size effects.  A recent review on this topic
and extensions thereof can be found in Ref.~\cite{Forquin}.

Alternative approaches based on a purely deterministic picture was proposed by
Francfort and Marigo\cite{Francfort}, as an extension of Griffith's theory.
This theory, although appealing, raises subtle mathematical issues ---
which may in fine resume to physical questions --- related to how small
scales may or may not disappear at the continuum limit for such singular
problems.  Another direction of attack consists of introducing a more
complicated picture for the crack tip, through a ``fracture process zone''
whereby, at microscopic scales, a displacement discontinuity can still
allow for the transmission of stress.  Such models following the initial
work of Barenblatt~\cite{Barenblatt} illustrate the fact that initiation
and propagation can be incorporated in the same picture if in addition
to toughness, a critical stress is also considered. This is yet another way
to stress that some microscopic length scale will survive in this macroscopic
picture.  Leguillon~\cite{Leguillon} for instance proposes such a dual
criterion for accounting for crack initiation in the spirit of these
fracture process zone models.  Although this latter class of theories
emphasizes the importance of small scales, materials are always treated
as if they were homogeneous.

Thus, from a continuum mechanics point of view, the choice is left between
approaches that either emphasize the geometry of defects (without mechanical
manifestation other than linear elasticity) or focus on fine details of
local variations of mechanical free energy (not to say damage) but
ignores material variability.

It is therefore interesting to consider discrete models (where no
ambiguity is left about microscopic limits) and where disorder is explicitly
considered.  On the one hand, such models may be considered as too
simplistic for portraying any specific materials, however, on the other
hand, they may shed some light on statistical size effects that may be
expected. This is very important as scaling is not neutral. \revision{If interpolation based on Weibull distribution is fair when the same sizes and level of probabilities as determined in identification}, it becomes very fragile when extrapolation is needed.

Hans Herrmann\cite{deArcangelis,Kahng,Herrmann} has been one of the
pioneers in this matter with first papers on electrical analog, ``fuse
networks'', to brittle material published in 1985.  However, a direct
numerical simulation reveals to be a difficult pathway to reaching
definitive conclusions, \revision{as results are prone to slow
  corrections to scaling, and become expensive in computational power
  as one moves to two or three dimensions. \cite{Barai-PRE13}}
Nevertheless, those simulations could demonstrate quite clearly that
the weakest link approach could not be applied at the smallest scales.
Therefore, it was already clear from the start that neither approach
based on prior defect statistics only nor on a more sophisticated
interaction model but within a homogeneous medium description were
applicable, and hence the merit of the statistical laws and the
attached corresponding size effects could be questioned.

To progress along these lines, fiber bundle models~\cite{Peirce,Daniels}
play a key role. Indeed despite their simple definition, and their
introduction about 90 years ago, an exhaustive characterization of their
behavior has only been achieved in some particular
cases\cite{Hemmer,Pradhan-RMP10,HHP15}.

These models consist of a collection of elastic brittle fibers with random
strength loaded in parallel.  One very important feature of these models
leading to different scaling regimes is the way load is distributed
amongst the unbroken fibers.  In the ``Equal Load Sharing'' (or ELS),
\cite{Daniels} load is equally distributed over all surviving fibers.
This corresponds to clamping all fibers onto a rigid substrate onto which
the external load is applied.  Because of this even distribution, the
dimensionality of the substrate plays no role.  This limit can thus
be seen as a mean-field model which is analytically tractable.

The opposite limit of Local Load Sharing, (or LLS), is based on a
local redistribution of the load carried by a fiber onto its unbroken
neighboring fibers \cite{HHP15,Harlow}.  Here the breakdown scenario is very
different from ELS, in the sense that after a first diffuse damage
regime, (where fibers breakage is spread over the entire system), a
localized region hosts a crack which --- after its nucleation --- grows without limits.  Macroscopically, this failure appears as brittle but after a
significant number of fibers have been broken.
However, the LLS model suffers from some ambiguity on the way the
load is redistributed after fiber breakage.  This ambiguity becomes
even more pronounced in two dimensions and more. In one dimension, some
aspects of this model are analytically tractable \cite{HHP15,KHH97}. However,
mostly, numerical methods must be used.

There are also other more realistic variants of the fiber bundle model.
The Soft Clamp Model \cite{DRP99,DRP99b,BHS02,SGH12} has the fibers attached to clamps that respond elastically. The clamps are infinite half spaces and the force distribution among the fibers are due to the elastic response of the clamps.

\revision{The transition between ELS and LLS has also been studied by varying continuously the sharing rule trough a simple power law stress redistribution function $\sigma \sim r^{-\gamma}$, where $r$ is the distance from the broken fiber \cite{Hidalgo-PRE02,Yewande-PRE03,Hidalgo-PA05}. It was shown that the stress-transfer function decay exponent dramatically impacts breaking processes, ultimate strength and creep regime. In the limiting case of ELS, the breaking of fibers is a completely random nucleation process while, for LLS, it becomes a nucleation problem.}

\revision{In the present study, we modify the Soft Clamp Model by replacing the infinite half space clamps by thin elastic membranes. This corresponds to the LLS limit, that is to say an infinite exponent for the stress-transfer function ($\gamma \rightarrow \infty$) in \cite{Hidalgo-PRE02,Yewande-PRE03,Hidalgo-PA05}. We note that both in the
Soft Clamp model and the Membrane model, one of the clamps can be replaced by an infinitely rigid one.  For symmetry reasons, this model will have
the same behavior as were the two clamps equal.}

\revision{Other physically realistic models or limits could have been
  considered.  In particular, an elastic plate described through
  Love-Kirchhoff theory has naturally a bending stiffness.  Indeed,
  there is no objection to considering such a model, (and it has
  already been considered for instance in \cite{DRP01}) however, in 1D
  where the plate reduced to an elastic beam, it does not coincide
  with the LLS model precisely because of the flexural effect and
  continuity of the rotation.  This introduced a dampened oscillatory
  behavior close to a crack.  Because our aim is here to introduce a
  natural extension of LLS model in 2D that suffers no ambiguity in
  its definition, we will stick to the definition of the Soft Membrane
  limit.}

\revision{Let us also mention a related model, namely that of {\it
    shear} delamination of a thin film from a rigid substrate as
  introduced by Zaiser {\it et
    al.}\cite{Zaiser-JSM09a,Zaiser-JSM09b,Zaiser-JSM11}.  A modeling
  proposed in \cite{Zaiser-JSM09a} shows that the in-plane
  displacement obeys an anisotropic Laplace equation, that can be
  somehow compared to the Poisson equation ruling the out-of-plane
  displacement for the membrane under distributed load.  The source
  term that is present in the Poisson equation is the result of the
  external loading and the balance equation leading to damage and
  crack propagation.  A similar effect is naturally present in other
  modelings~\cite{DRP99,DRP99b,DRP01,Zapperi-EPJB00}.  Since it
  controls the instability of crack propagation which is the hallmark
  of the LLS model, it is the analogy with models that have no
  distributed loadings that should be considered carefully.}
%

\revision{In the present study, we show that it is possible to define and compute a toughness allowing for understanding the post-peak behavior, i.e. after the maximum load has been reached.} Moreover, this limit is also shown to match that of a crack front pinning model.  The simplicity of the model finally allows us to study its behavior in the continuum limit via a perturbative approach, drawing promising path from nucleation to crack propagation in brittle heterogeneous materials.

 \begin{figure}[ht!]
 \begin{center}
 \includegraphics[width=0.90\textwidth]{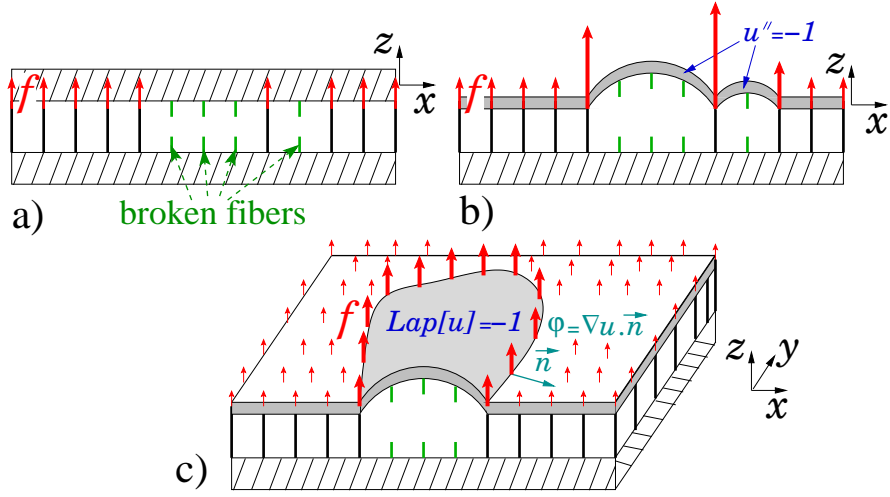}
 \end{center}
 \caption{\label{fig:LoadSharingRule}
Load sharing rules in a fiber bundle between clamps that are elastic
membranes.  The lower membrane is infinitely stiff, whereas the upper
membrane responds elastically.
A macroscopic load $F$ is applied to the system through a homogeneous pressure onto the membrane and is
shared between unbroken fibers subjected to local forces $f$.
a) When the upper membrane  is stiff, the model behaves as the ELS model.
b) In one dimension, a very soft upper membrane (or rather, string) will
cause the model to behave as the LLS model.
c) In two dimensions, a very soft upper membrane \revision{will transfer an additional load only to fibers adjacent to already broken ones.}  }
 \end{figure}

\section{Soft Membrane Fiber Bundle in One Dimension}

The fiber bundle initially consists of $N$ fibers distributed
homogeneously over a linear domain of size $L=N$, at positions labeled
by integer $x$ coordinates. To reduce the importance of edge effects,
periodic boundary conditions are implemented. Each fiber is assumed to be
elastic-brittle.  Initially, all fibers are present. Then, through the
application of load, fibers will be broken one at a time.  A sequence index, $t$, is introduced which simply counts the number of
broken fibers. \revision{For simplicity, $t$ will be referred to as ``time''.}
Let us emphasize that the model itself does not depend on the physical time.
In the elastic regime, a macroscopic load $F$ is applied in the system,
and is shared between unbroken fibers as illustrated in
Fig.~\ref{fig:LoadSharingRule}.
A fiber at position $x$ is subjected to a local force $f(x,t)$. If
the fiber is broken, then $f(x,t)=0$.  If not, because of
elasticity, $f(x,t)$ is proportional to the external load,
$f(x,t)=F\varphi(x,t)$, where $\varphi(x,t)$ characterizes the
load redistribution on the fiber, and that depends on the distribution
of broken fibers and hence on $t$.  Each fiber is characterized by its
strength, $f_c(x)$, independent of time.  At each instant,
the macroscopic breaking load is the value of $F$ such that one and only one
fiber reaches its threshold, $f(x,t)=f_c(x)$, hence
    \be
    F_c(t)=\min_x\frac{f_c(x)}{\varphi(x,t)},
    \ee
\revision{which corresponds to an extremal dynamics.}

The load sharing rule is what allows us to define $\varphi(x,t)$.  For
instance, ELS means that all surviving fiber support the same
force as shown in Fig.~\ref{fig:LoadSharingRule}a. Overall balance requires
that $\sum \varphi(x,t)=1$, and hence $\varphi(x,t)=1/(N-t)$ if the fiber
is unbroken, and 0 else.

The membrane model is equivalent to the LLS rule in 1D when
the membrane (string) is very soft. We now discuss the LLS rule in some
detail. As above mentioned that balance imposes $\sum \varphi(x,t)=1$.
In 1D, it is rather straightforward to arrive at the fact that
for any interval of $m$ consecutive fibers broken, an additional
force of $m/2N$ is transfered to the two surviving fibers nearest
neighbor to the end of this interval. This comes in addition to
the $1/N$ force that it always present. Note that a fiber may be
neighbor of two intervals of $m_l$ and $m_r$ failed bonds
respectively for left and right, and hence its load partition
coefficient will be
    \be
    \varphi(x,t)=\frac{2+m_l+m_r}{2N}
\label{LLS}
    \ee

The Soft Membrane model is now considered.  In this case, the
deflection $u(x)$ of this string under the applied loading has to be
solved.  If we consider that the fibers are much stiffer than the
string, if the fiber is present at position $x$, then $u(x)=0$.
Otherwise,
    \be
    \Lap[u](x)=-1
    \ee
where $\Lap$ is the discrete laplacian $\Lap[u](x)=u(x+1)-2u(x)+u(x-1)$. The load transferred by a membrane to its boundary writes $f^T = -{\mathbf
\nabla} u\cdot {\mathbf n}$ where ${\mathbf n}$ is an outer unit vector, normal to the membrane boundary (see Fig. \ref{fig:LoadSharingRule}). In 1D, this means that a fiber located at $x$ receives an additional transfer $f^T(x)=[u(x+1)-u(x)]+ [u(x-1)-u(x)]$ where we separated the contributions of the left and right neighbors. In this case, the load applied to a surviving
fiber $x$ (which is such that $u(x)=0)$ is
    \be
    \varphi(x,t)=(1/N)\left(1+\sum_{y~n.n. x} u(y)\right)
    \ee
where a summation over nearest neighbors ($n.n.$) of $x$ is involved.
The occurrence of the second order differential operator $\Lap$ comes
from a conservation (balance) equation so that the total load applied
onto the system is finally transferred to the unbroken fibers. The
deflection $u(x)$ of a membrane of extremities $x_i$ and $x_f$
writes $u(x) = (x-x_i)(x_f-x)/2$ and the load transfers
$f^T(x_i)=f^T(x_f)=(x_f-x_i-1)/2$. For a fiber surrounded by two
intervals of $m_l$ and $m_r$ bonds, one thus recovers the very same
load partition coefficient than in the 1D LLS fiber model given
above (Eq. (\ref{LLS}).

\begin{figure}[ht!]
 \begin{center}
 \includegraphics[width=0.45\textwidth]{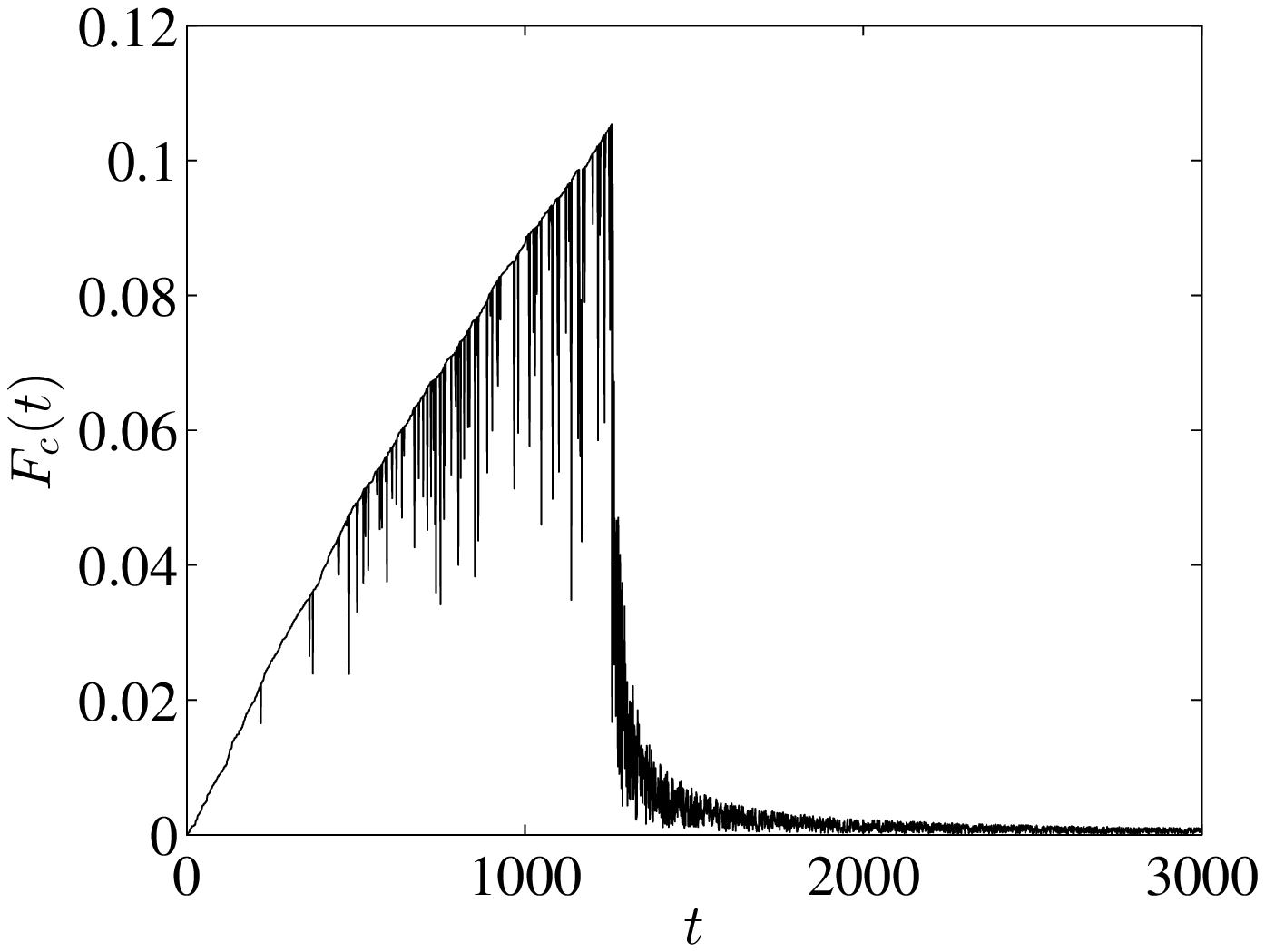}
 \includegraphics[width=0.45\textwidth]{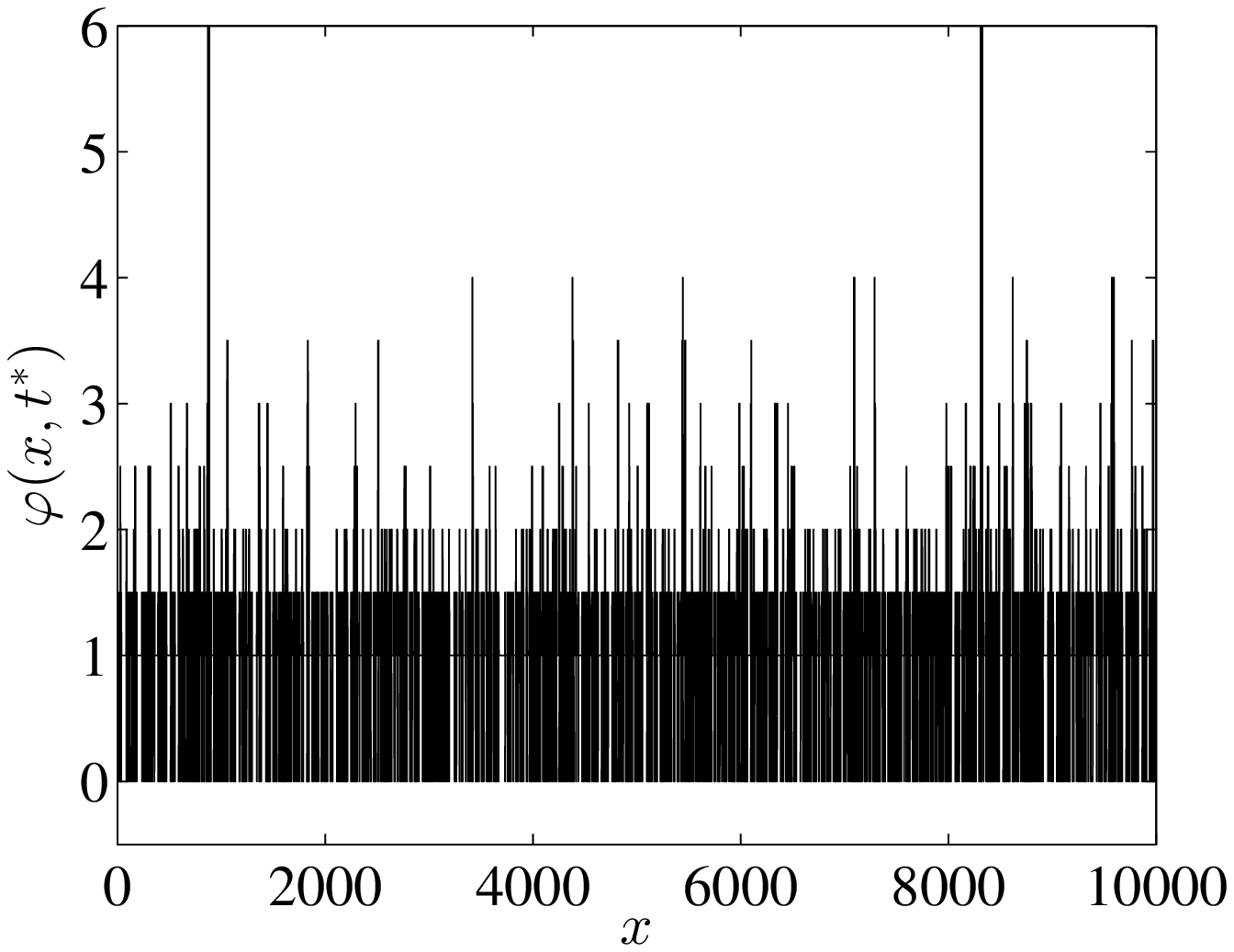}
 \end{center}
 \caption{\label{fig:1DLoad} (a) Load versus time in 1D ($N=10000$ and
   periodic boundary conditions). The critical time is the one where
   the maximum force is reached, $t^*=1256$.  (b) Local force
   distribution $\varphi(x,t^*)$ at the critical load.}
 \end{figure}

Figure~\ref{fig:1DLoad}a shows the classical 1D model load versus time $t$.
We considered here a uniform distribution of fiber strength in $[0;1]$.
The critical load is reached at $t=t^*=1256$.
Thus, the weakest link
assumption is indeed far from being applicable.  Figure~\ref{fig:1DLoad}b
shows the redistribution function at the peak force.  We see that the
largest fiber load $\varphi$ comes from an interval of 10 consecutive
broken fibers.

\section{Soft Membrane Fiber Bundle in Two and More Dimensions}

The generalization of the 1D soft membrane model to two dimensions (2D)
is straight forward. The system is a $L\times L$ square, so that
the number of fibers is $N=L^2$.  A fiber location is designated either
by a 2D vector $\bm x$ or by its integer coordinates $(x,y)$.  The question
is now to define the redistribution function.

The membrane deflection, $u(\bm x)$, obeys
    \be\left\{
        \ba{ll}
        u(\bm x)&=0\\
        \Lap[u](\bm x)&=-1
        \ea \right.
    \textrm{if fiber at }\bm x\textrm{ is } \left\{
        \ba{l}
        \mathrm{intact}\\
        \mathrm{broken}
        \ea\right.
    \ee
where $\Lap[u]$ is again a finite difference laplacian operator
    \be
    \Lap[u]=\sum_{\bm y~n.n. \bm x} (u(\bm y)-u(\bm x))
    \ee

The redistribution function on a surviving fiber in $x$ is simply
    \be
    \varphi(\bm x,t)=(1/N)\left(1+\sum_{\bm y~n.n. \bm x} u(\bm y)\right)
    \ee
i.e., precisely the same expression as in the 1D model.

\begin{figure}[ht!]
 \begin{center}
 \includegraphics[width=0.45\textwidth] {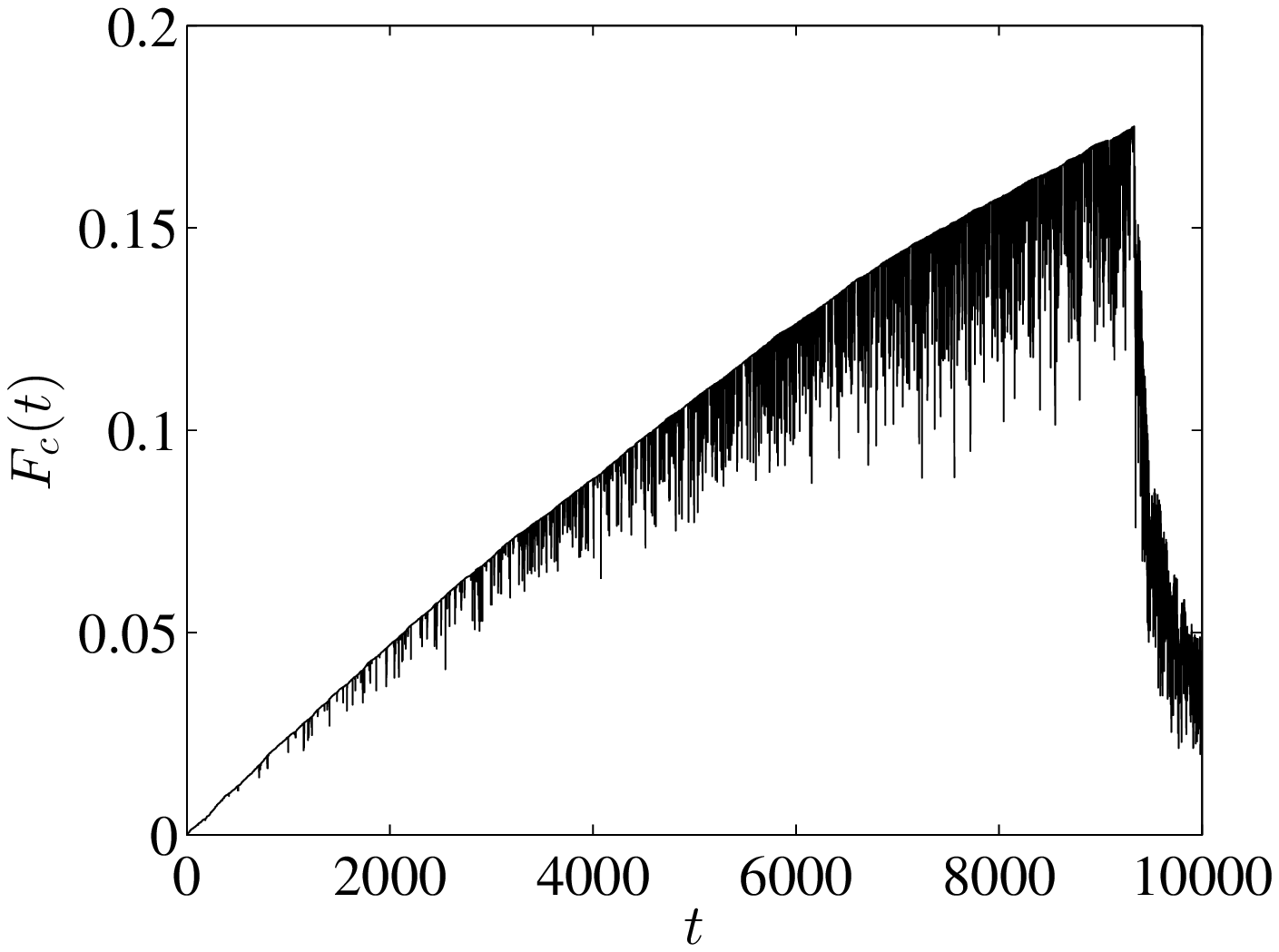}
 \includegraphics[width=0.45\textwidth] {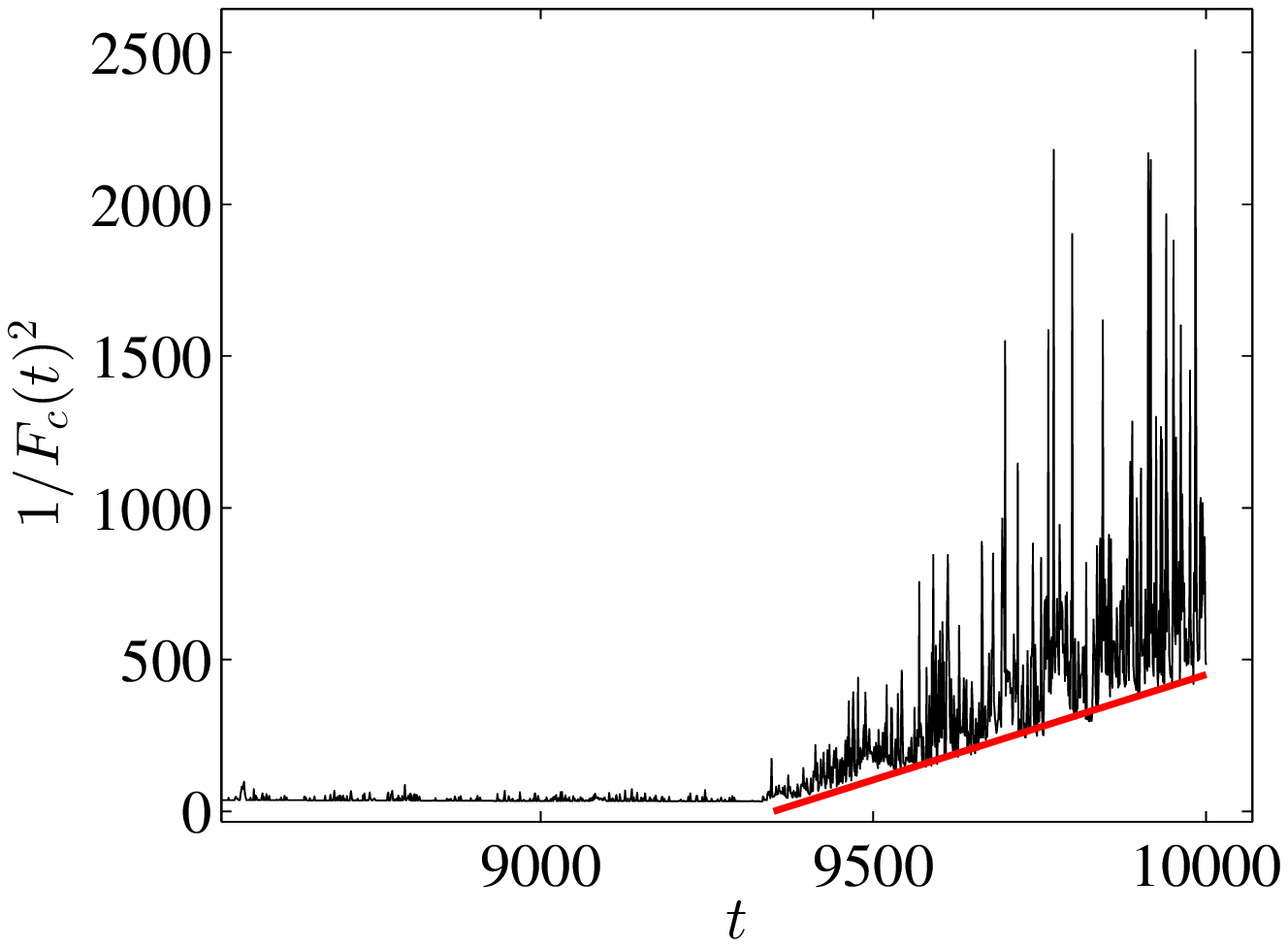}
 \end{center}
 \caption{\label{fig:2DLoad} (a) Load versus time for the Soft
     Membrane fiber bundle model in 2D.  (b) Post-critical analysis
   of critical load versus time.  The red line belongs to the circular
   crack analysis.}
 \end{figure}

\begin{figure}[ht!]
 \begin{center}
 \includegraphics[width=0.45\textwidth]{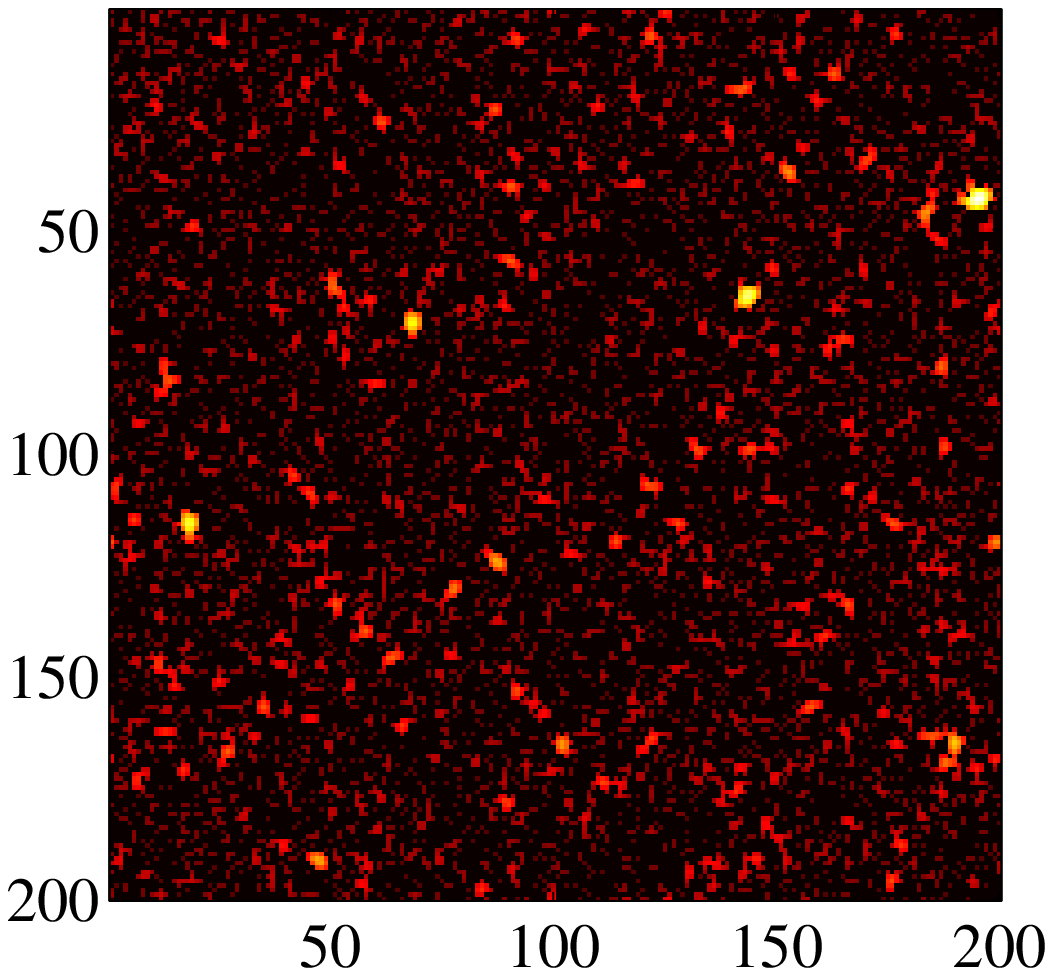}
 \includegraphics[width=0.45\textwidth]{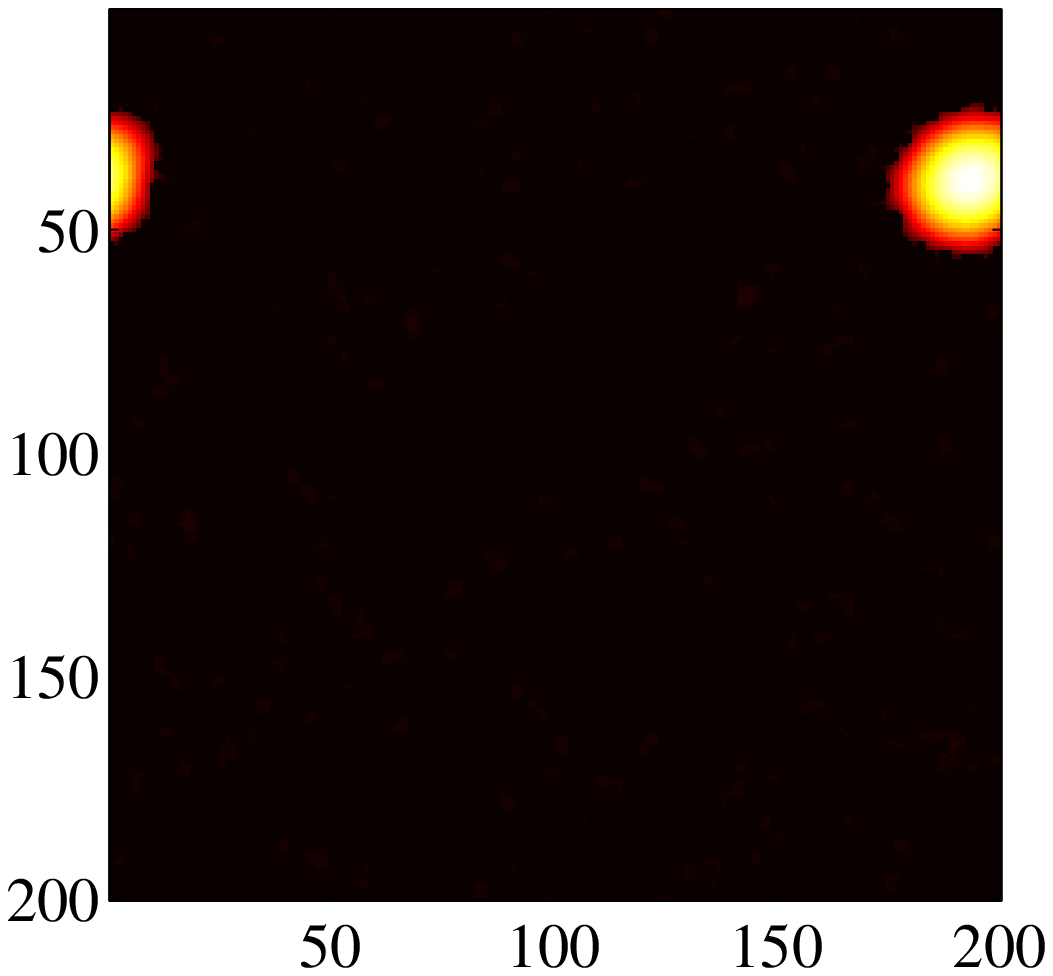}
 \end{center}
 \caption{\label{fig:2D_u} Membrane displacement $u(\bm x)$ in the 2D Soft Membrane
fiber bundle at the critical load (left), and at a later stage (right).
The center of the critical defect can be guessed from the late stage.
The periodic boundary conditions are also visible.}
 \end{figure}

\begin{figure}[ht!]
\begin{center}
 \includegraphics[width=0.45\textwidth] {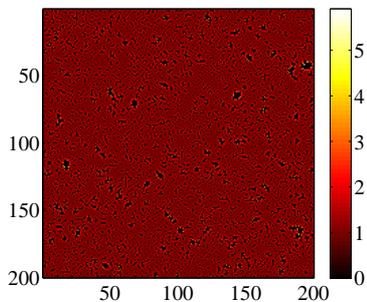}
\end{center}
 \caption{\label{fig:2Dfloc} Local redistribution function
$\varphi(\bm x,t^*)$ at the critical load.}
\end{figure}

Figures~\ref{fig:2DLoad}, \ref{fig:2D_u} and \ref{fig:2Dfloc}
illustrate the load versus time, the displacement and local force per fiber
at the critical state, and the failure time map at a late stage.  It is
apparent that, as expected, a critical defect is nucleated and finally
grows with no possible arrest.  This growth takes the form of a circular
crack as seen in Fig.~\ref{fig:2D_u}. 
Let us stress again that the  fraction of broken fibers at the peak
stress is a rather large proportion, $t^*/N\approx 23\%$.  Again the
weakest link approach is inapplicable.  Simultaneously, when
considering the overall macroscopic force versus time,
Fig.~\ref{fig:2DLoad}a, there are no fore-signs of nucleation of the
critical defect.

This observation can be rationalized, in the spirit of homogenization, by
considering a uniform equivalent threshold, $f_c^{eq}$.
Considering a circular cluster of broken bonds,
the load supported by the broken
fibers $\pi R^2 F$ is distributed over the edge, leading to a force per
fiber of  $R(t) F/2$ and hence $F_c(t)=2f_c^{eq}/R(t)$.  The radius
is related to time as $\pi R(t)^2=t-t_0$. Thus
    \be
    F_c(t)=\frac{2\sqrt{\pi}f_c^{eq}}{\sqrt{t-t_0}} \label{effective-strength}
    \ee
Figure~\ref{fig:2DLoad}b shows that this analysis holds after
nucleation of the critical defect, as $1/F_c(t)^2$ has indeed a lower
envelope obeying an affine law.  The straight line fit allows to define
an effective threshold equal to $f_c^{eq}\approx 0.34$.



\section{Measurement of Toughness}

Having understood the late scenario of a crack propagation, one may study
it directly: a specific crack pattern already present in the initial
state can be designed.  For instance, half of the domain
(say $0\le y < L/2$) consists of broken fibers in the initial state.
The problem is thus initially similar to the 1D case.  The membrane
deflexion field is
    \be
    u(x,y)=(1/2) y(L/2-y)
    \ee
and hence the force acting along the edges is the total load $(1/2)L^2F$
divided by the front length (on both edges) $2L$ or $FL/4$.  It can
also be computed as $\Lap[u]$.  In our case, for $L=200$, the force
acting on the front edge is 50 times higher than in the bulk.  Thus
most (but not all) of the activity is taking place at the front. Note,
however, that as the front propagates as a whole, the overload at the front
increases, and hence the diffuse damage regime is only taking place in an
initial transient regime, which vanishes as $L\to \infty$.

\begin{figure}[ht!]
 \begin{center}
 \includegraphics[width=0.45\textwidth] {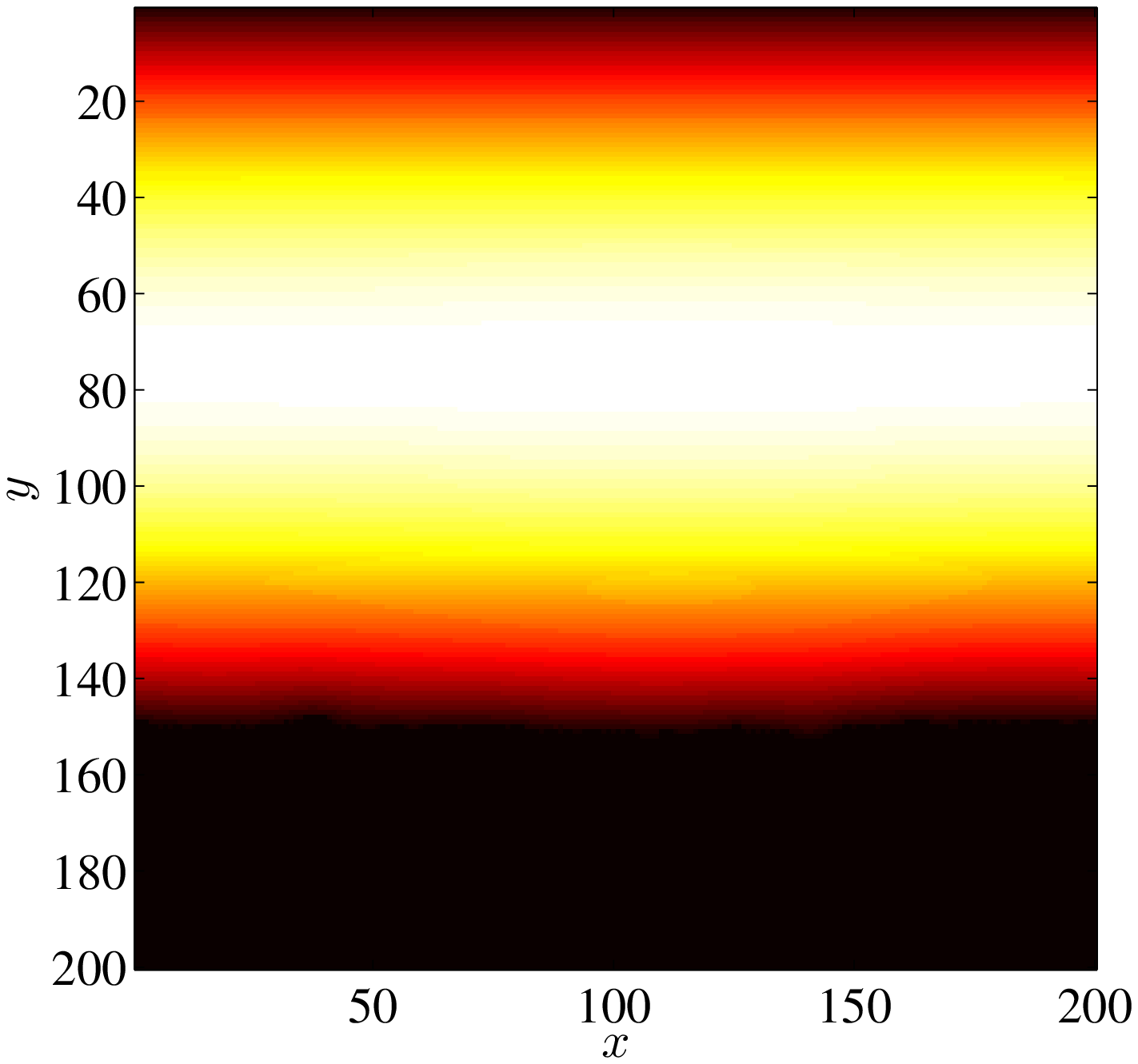}
 \includegraphics[width=0.45\textwidth] {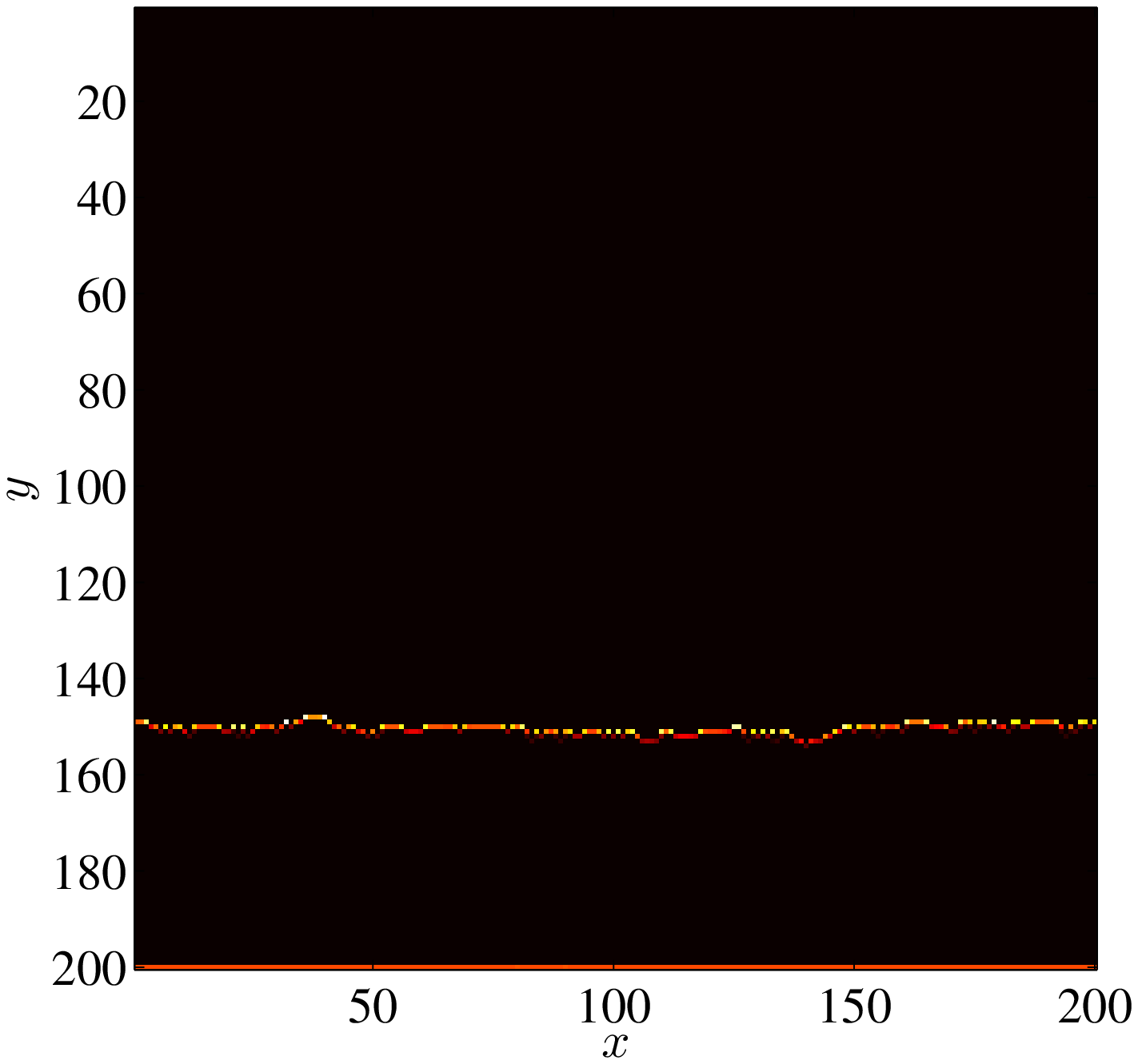}
 \end{center}
 \caption{\label{fig:2DfrontUf} Displacement $u(x,y)$ (left) and
local force $f(x,y)$ (right) after some progression.  }
 \end{figure}

\begin{figure}[ht!]
 \begin{center}
 \includegraphics[width=0.8\textwidth] {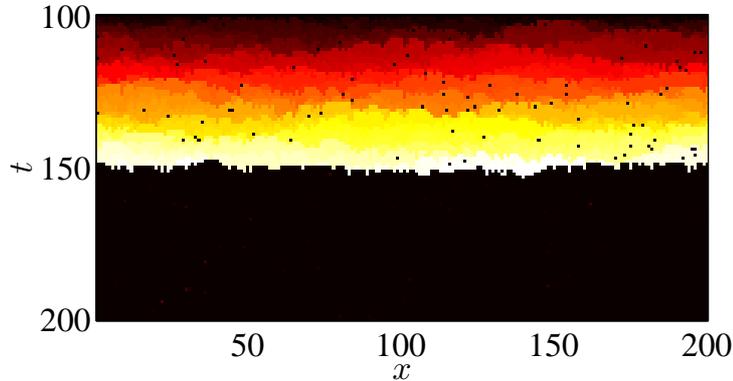}
 \end{center}
 \caption{\label{fig:2Dfront_t} Map of failure time where the downwards front
progression  is easily observed.}
 \end{figure}

Periodic boundary conditions being implemented, one may stop one of the
two fronts ($y=L$ in our case) by inserting a line of fibers
with very high strength.  Figure~\ref{fig:2DfrontUf} shows the
displacement field and force acting on the fibers after some
progression starting from the previous initial conditions.  It
is observed that a slightly roughened front progress from
$y=L/2$ onward.  Figure~\ref{fig:2Dfront_t} shows the map of the
time at which the fibers were broken.  One can easily observe indeed
the above mentioned front propagation, together with some ``dust''
which corresponds to the early damage of bonds being broken ahead of
the front (in the present case, all bonds having a strength less
than $2/L\sim 0.01$ are broken in a first stage). \revision{Note that the dynamics of the
preexisting crack does not proceed by avalanches since we perform extremal dynamic simulations, i.e.
each ``time'' step only one fiber is broken.  }

\begin{figure}[ht!]
 \begin{center}
 \includegraphics[width=0.45\textwidth] {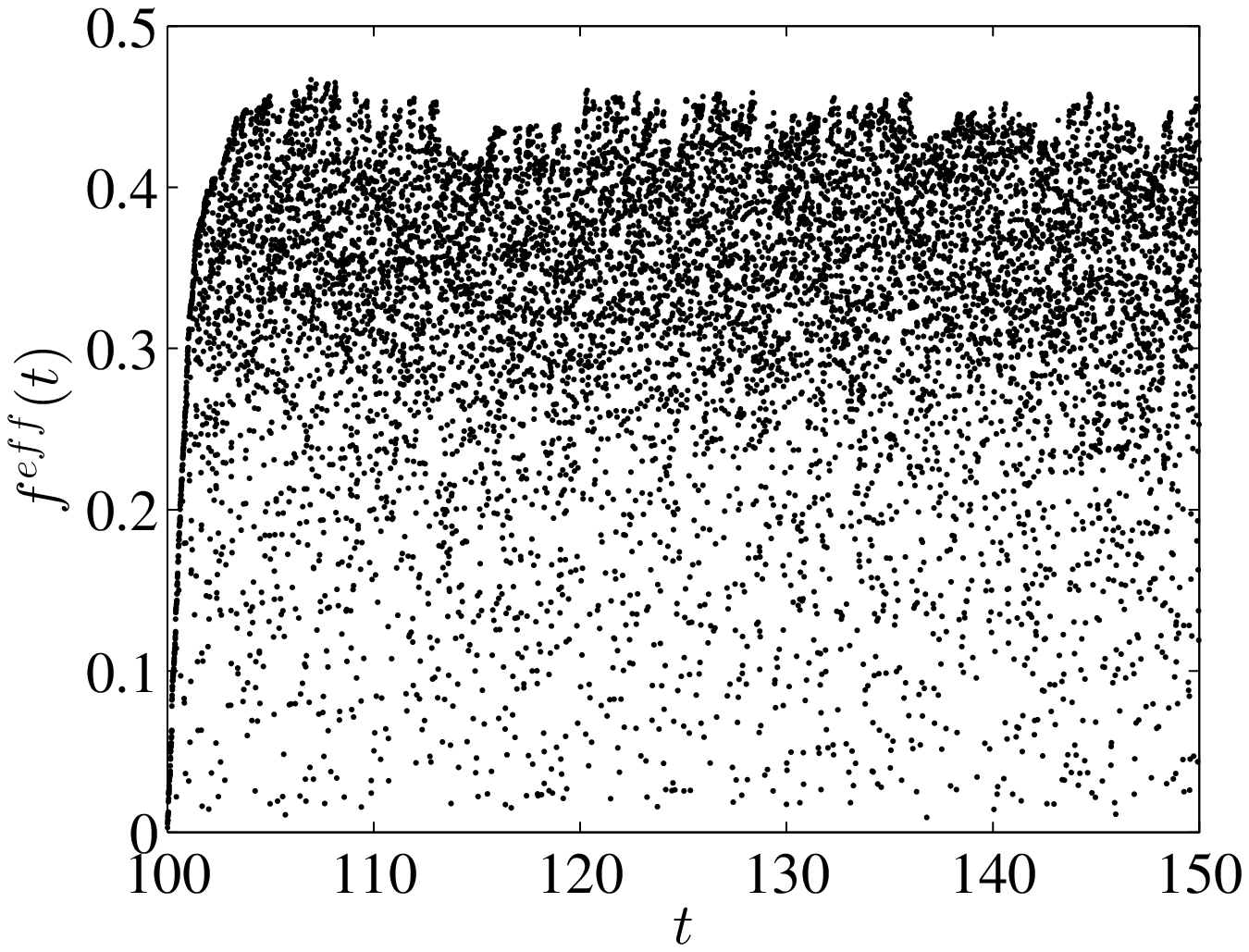}
 \includegraphics[width=0.45\textwidth] {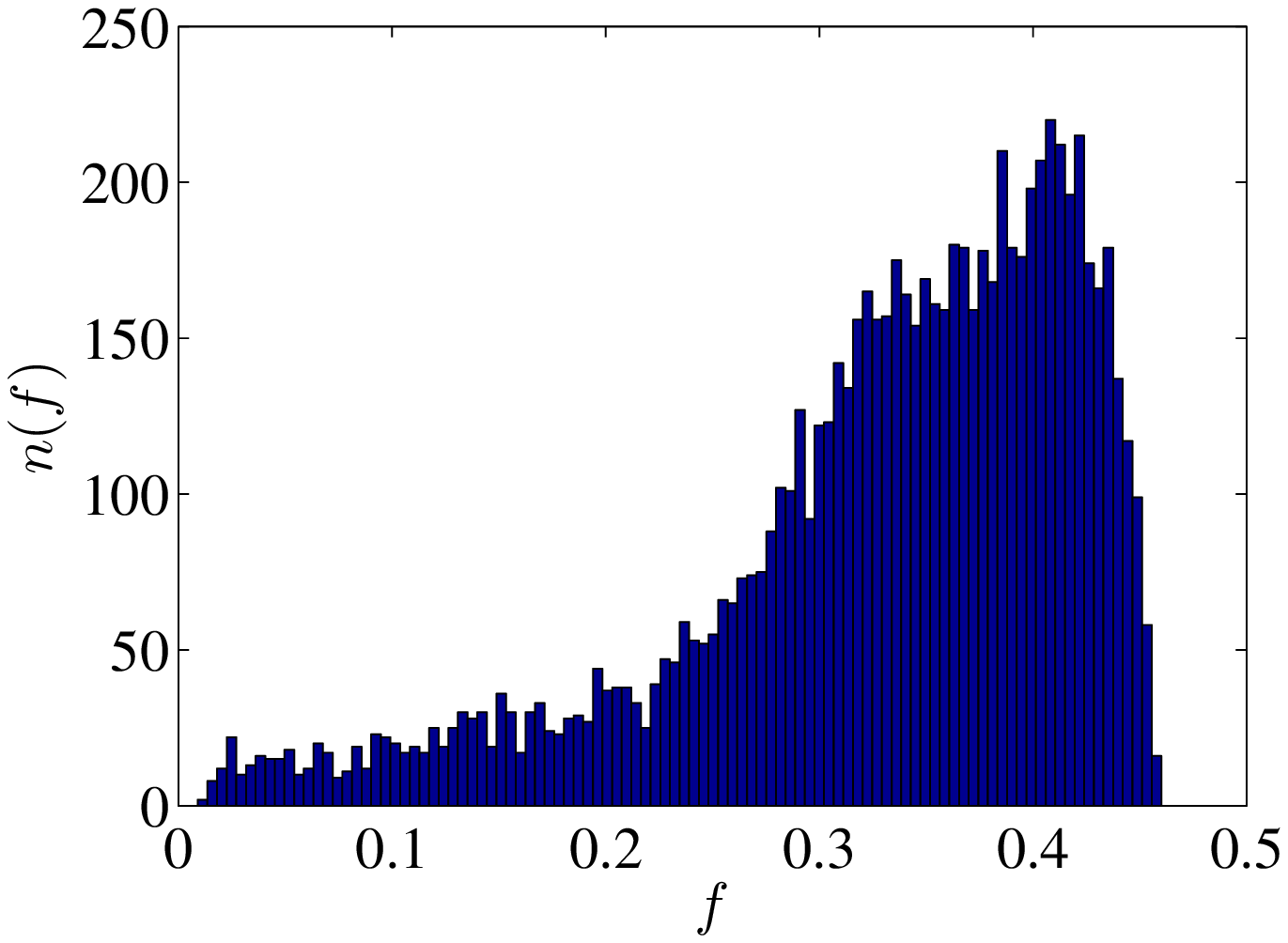}
 \end{center}
 \caption{\label{fig:2DfrontFc} (a) Effective fiber strength $f_c^{eff}$,
deduced from the critical load by a multiplication of half the mean
crack length. (b) histogram of the instantaneous critical load,
where the asymptotic toughness is the upper bound of this distribution.}
 \end{figure}

Figure~\ref{fig:2DfrontFc} shows the load versus time.  One observes
again clearly the first initial (diffuse) damage regime where fibers
are not located at the front.  The interesting regime is the front
progression stage.  In the continuum, if the fiber strength
$f_c^{eff}$ were uniform, one would expect a uniform progression
$y=y_0+t/L$ (where $y_0=L/2$ in our case), so that the force
exerted at the front would be $(y_0+t/L)F/2$ and hence the
critical load would be
    \be
    F_c(t)=\frac{2f_c^{eff}}{(y_0+t/L)}
    \ee
Hence one can define an (instantaneous) effective bond strength,
$f_c^{eff}$, from the critical force through $ F_c(t)(y_0+t/L)/2$.
This quantity is shown in Figure~\ref{fig:2DfrontFc}.  The
maximum value of the latter strength appears as a well-defined
quantity which is the macroscopic effective strength.
The histogram shown in Fig.~\ref{fig:2DfrontFc} is another way of
revealing the threshold.  It allows us to estimate
    \be\label{eq:toughness}
    f_c^{eff}\approx 0.46
    \ee
as compared to the elementary (mean field) approximation of
$\langle f_c\rangle=0.5$ that ignores the effect of load redistribution.

It is noteworthy that this estimate is also slightly larger than the
one estimated from the random nucleation case.  This may be due to
the fact that a somewhat larger amount of damage has been
experienced in the plain 2D lattice at the onset of nucleation.
Estimating this damage at this point as roughly $D=0.23$ ($D$ is
the fraction of missing bonds), leads to an effective
decrease of the effective strength (when averaged over the front)
down to $(1-D)f_c^{eff}\approx 0.34$ very consistent
with the previous estimate. To be complete, one should also correct
the above estimate by the damage level of the front type geometry,
but because of the large system size, the damage is roughly $D=0.01$,
and does not affect our estimate.

The result in Eq.~(\ref{eq:toughness}) is far from trivial.  We may do the same
same analysis in 1D where, in contrast, the result is trivial,
$\langle f^{eff}_c\rangle$ is the average of the threshold distribution.
This is so since the fibers break sequentially as a zipper is opened.

\section{Continuum limit: Depinning Model of Crack Propagation}

We now consider the case of a penny-shape crack in the continuum
limit. Accounting for the front roughness induced by the fiber
strength disorder enables to rationalize the propagation stage of the
discrete model.
Let us consider a (rough) penny-shaped crack of equation
\be
\rho = R(\theta) = R_0 \left[1 + \varepsilon h(\theta) \right]
\ee
in polar coordinates, where $\varepsilon$ sets the amplitude of the
radius fluctuations.
The membrane deflection $u(\rho,\theta)$ obeys
\be\ba{rlcl}
\nabla^2 u &= -1  &\qquad{\rm if} \qquad   &\qquad\rho < R(\theta)\\
u &= 0  &\qquad{\rm if}     \qquad &\qquad\rho = R(\theta)
\ea\ee
At zeroth order, for $\rho\le R_0$, the solution for a circular crack of radius $R_0$ writes
    \be
    u_0(\rho,\theta) = \frac{1}{4} \left( R_0^2 - \rho^2\right)
    = \frac{1}{4} \left( R_0^2 - z \overline{z} \right)
    \ee
where we defined the complex variable $z=\rho e^{i\theta}$ and
its conjugate $\overline{z}=\rho e^{-i\theta}$.
Let us now write $u = u_0 + \varepsilon R_0^2 u_1$.
Since $\nabla^2 u_0 = -1$ and $u_0(R_0,\theta)=0$,
the first order contribution to the membrane deflection $u_1(\rho,\theta)$ obeys
    \be\ba{rlcl}
    \nabla^2  u_1 &= 0 &\qquad{\rm if}\qquad &\qquad\rho < R(\theta)\\
    \varepsilon R^2 u_1
    (\rho,\theta)&= -u_0(\rho,\theta)  &\qquad{\rm if} \qquad&\qquad\rho = R(\theta)
    \ea\ee
The boundary condition can be rewritten explicitly as
\be
\varepsilon R^2 u_1 \left[  R_0 \left(1 + \varepsilon h(\theta) \right)\right]
=-u_0\left[  R_0 \left(1 + \varepsilon h(\theta) \right)\right]
\ee
so that, at first order
\be\label{BC1}
u_1(R(\theta),\theta) = \frac{1}{2} h(\theta)\;.
\ee
In order to evaluate $u_1$ inside the crack, $\rho<R(\theta)$, it is written as a Laurent series (polynomial expansion in $z$):
\be
u_1(z,\overline{z}) = \alpha_0+\sum_{n=1}^{N/2} \alpha_n z^n +
\sum_{n=1}^{N/2} \overline{\alpha_n}\; \overline{z}^n
\ee
In complex variables the Laplacian operator is $\Delta = 4 \partial_z\partial_{\overline{z}}$ so that this expansion immediately satisfies the equation $\Delta  u_1 =0$ (in other words, $u_1$ being the real part of an holomorphic function, it is harmonic).  The restriction of this function on the circle $\rho = R_0$
can be identified with a Fourier series and gives an immediate
solution to the boundary condition,  Eq.~(\ref{BC1})
\be\ba{rlll}
 u_1 (R_0,\theta)&= \alpha_0 
&\displaystyle{+\sum_{n=1}^{N/2} {\alpha_n}\; {R_0}^n e^{in\theta}}
&\displaystyle{+\sum_{n=1}^{N/2} \overline{\alpha_n}\; {R_0}^n e^{-in\theta}}\\[10pt]
 h(\theta) &= h_0 &\displaystyle{+
\sum_{n=1}^{N/2} {h_n}\;  e^{in\theta} }
&\displaystyle{+\sum_{n=1}^{N/2}\overline{h_n}\;  e^{-in\theta} }
\ea\ee

The first order membrane deflection
$u_1(\rho,\theta)$ thus reads:
\be
2u_1 (\rho,\theta)=h_0
+\sum_{n=1}^{N/2} \left(\frac{\rho}{R_0}\right)^n{h_n}\;
e^{in\theta}
+\sum_{n=1}^{N/2} \left(\frac{\rho}{R_0}\right)^n
\overline{h_n}\;  e^{-in\theta}
\ee
The force $\varphi= -{\mathbf \nabla}u \cdot {\mathbf n}$ induced at the boundary can now be computed at first order
    \be\ba{rll}
    {\mathbf n}(\theta)
    &=&{\mathbf e}_\rho -\varepsilon R_0 h'(\theta)\;{\mathbf e}_\theta\\[10pt]
    {\mathbf \nabla}u(\theta)
    &=&\left[ \frac{\partial u_0 }{\partial \rho}(R_0)
    +\varepsilon R_0 h(\theta) \frac{\partial^2 u_0 }{\partial \rho^2} (R_0)
    + \varepsilon R_0^2 \frac{\partial u_1 }{\partial \rho}
    (R_0,\theta) \right]{\mathbf e}_\rho \\[10pt]
    &&+\left[ \varepsilon \frac{\partial u_0 }{\partial \rho}(R_0) h'(\theta)
    +\varepsilon R_0 \frac{\partial u_1 }{\partial \theta} (R_0,\theta)\right]{\mathbf e}_\theta\\[10pt]
    %
    \varphi\left[R_0(1+\varepsilon h(\theta))\right]
    &=&(R_0/2) \left[1+ \varepsilon h(\theta)
    -\varepsilon  f_1 (\left\{h\right\},R_0,\theta)\right]
    \ea\ee
where ${\mathbf e}_\rho$ and ${\mathbf e}_\theta$ are
the unit vectors in polar coordinates and
    \be
    f_1 (\left\{h\right\},R_0,\theta)
    =\sum_{n=-N/2}^{N/2} |n|{h_n}\;  e^{in\theta}
    \ee

This result is reminiscent of
the effect of a rough geometry on the stress intensity factor
of a mode I crack front as derived by Rice~\cite{Rice-JAM85}
in semi-infinite geometry, and more recently by Legrand {\it et al}~\cite{Legrand-IJF11} for a plate geometry:
    \be
    K_I \left[ x, z=h_0+ \varepsilon h(x)  \right]
    = K_I^0\left[1+\varepsilon\frac{1}{K_I^0}\frac{\partial K_I}{\partial z}
    h(x)-\varepsilon f(x)\right]
    \ee
with
    \be
    f(x) = A_1\sum_{n=-N/2}^{N/2} |n|{h_n}\;  e^{inx}
    \ee
where $A_1=1/2$ in the semi infinite case and $A_1=2$ in the plate
case. It is to be observed that the elastic interaction mediated by the membrane elasticity acts as a non-local line tension, stabilizing the crack front shape. \revision{This effect is quite generic, and was reported in the above references. It has also been derived by Zaiser {\it et al}~\cite{Zaiser-JSM09a} for the shear delamination problem.  It is also encountered in wetting problems\cite{Joanny-JCP84}.}

This line tension is consistent with the observation of the almost perfect circular shape of the crack observed in Fig.~\ref{fig:2D_u}, or the straight front of Fig.~\ref{fig:2DfrontUf}.  However, it is to be noted that the energy release rate increases as the radius of the penny shape crack grows.  Thus the overall expansion of the crack is unstable (under a constant applied pressure onto the membrane), but its shape tends to remain circular due to the line tension effect.  This can be read from the expression of the force redistribution in Fourier space:
\be
\varphi_n = \frac{R_0}{2} \left[\delta_n + \varepsilon (1-|n|)h_n \right]
\ee
The fact that $\varphi_{\pm 1}$ vanishes can be understood as the result of Galilean invariance as the first order mode in Fourier space corresponds to a mere translation of the crack which is expected not to have any effect on the crack driving force.
%
%

Crack propagation is thus controlled by the interplay between local strength disorder and long-range interface tension. One recognizes here the ingredient of a depinning problem, a paradigm that has been used to discuss various problems of front propagation in physics\cite{Kardar-PR98} and has been successfully applied to crack propagation in recent years\cite{Schmittbuhl-PRL95,SVR-IJMPC02,CVHR-JMPS04,DLV-PRL08,Bonamy-PR11,PVR-PRL13}.

It appears in particular in this framework that the crack front
roughness exhibits a scaling invariant character: when measured over a
length scale $\ell$ the width $w$ of the front scales as $w(\ell)
\propto \ell^\xi$ with an exponent
$\xi\approx0.4$\cite{Rosso-PRE02}. Moreover the effective force
$f(\ell)$ needed to depin a front of lateral extension $\ell$ follows
a universal distribution of width $\delta f(\ell) \propto
\ell^{-(1-\zeta)}$ and is bounded by a maximum value, the critical
force $f^*$\cite{SVR-IJMPC02,CVHR-JMPS04,VSR-PRE04}. The emergence of a critical
threshold and of the associated size effects is potentially of great
interest in the context of the discrete fiber bundle model developed
in the present study. The critical value $f^*$ indeed sets an upper
bound to the effective homogenized fiber strength defined in
Eq.~(\ref{effective-strength}). Moreover, the knowledge of the finite
size fluctuations of this effective strength for small defects should
allow for the study of the very early stages of crack propagation, say
in the spirit of the approach used in \cite{CVHR-JMPS04} for crack
arrest.

\section{Conclusion and Critical Perspective}
\label{Sec:conclusion}

We have formulated a soft membrane fiber bundle model that reduces to the
Local Load Sharing fiber bundle model in 1D.

Failure proceeds in two steps: There is a first regime of diffuse damage
which is followed by a second one
corresponding to the nucleation and growth of a penny shaped crack.
The latter can be characterized in terms of a mean toughness
(obtained from a simple combination of the strength $f_c^{eff}$ and
a length scale given by the fiber separation).  This toughness can
be measured by a direct crack propagation experiment.
This toughness is also altered by the mean damage state of the
material at the onset of brittle failure.

If much emphasis has been put on the damage regime and characterization
of the critical defect, it may be worth investigating the late
failure regime with classical tools. The above presented perturbation analysis
exploiting the Poisson equation, $\nabla^2 u=-1$,
relates this problem to a depinning crack propagation model formulated in the continuum with the same long range kernel relating the local energy release rate to the crack front morphology.
This may lead to a precise estimate of the effective toughness, but its main virtue is to shed some light on the fluctuations.  Indeed, this analogy leads to the prediction that the toughness fluctuation decays as a non-trivial power law of the front length.  
A challenge for the future is thus to derive a statistical law for the onset of crack nucleation in this soft membrane fiber bundle.

AH thanks P.C. Hemmer and S. Pradhan for countless discussions on this topic.

\end{document}